\documentclass[numberedappendix]{emulateapj}
\pdfoutput=1

\usepackage{epsfig, natbib}
\tightenlines

\countdef\decade=200
\advance\decade by \year
\countdef\hours=201	
\advance\hours by \time
\divide\hours by 60
\countdef\mins=202
\advance\mins by \hours
\multiply\mins by 60
\multiply\hours by 100
\countdef\miltime=203
\advance\miltime by \hours
\advance\miltime by \time
\advance\miltime by -\mins

\def\mathbi#1{\textbf{\em #1}}

\newcommand{\msun}{M_{\odot}}

\newcommand{\tff}{t_{\rm ff}}
\newcommand{\vecv}{\mathbi{v}}
\newcommand{\vecx}{\mathbi{x}}
\newcommand{\vecp}{\mathbi{p}}
\newcommand{\krho}{k_{\rho}}

\defcitealias{krumholz07a}{KKM07}
\defcitealias{krumholz08a}{KM08}

\begin{document}

\title{Metallicity and the universality of the IMF}

\slugcomment{Accepted to the Astrophysical Journal}

\shorttitle{Metallicity and the universality of the IMF}
\shortauthors{Myers et al.}

\author{
        Andrew T. Myers \altaffilmark{1},
        Mark R. Krumholz\altaffilmark{2},
        Richard I. Klein\altaffilmark{3, 4}, and
        Christopher F. McKee\altaffilmark{1,4}
        }

\altaffiltext{1}{Department of Physics, University of California, Berkeley,
Berkeley, CA 94720; atmyers@berkeley.edu}
\altaffiltext{2}{Department of Astronomy and Astrophysics,
         University of California, Santa Cruz, CA 95064}
\altaffiltext{3}{Lawrence Livermore National Laboratory, P.O. Box 808, L-23, Livermore, CA 94550}
\altaffiltext{4}{Department of Astronomy, University of California, Berkeley,
Berkeley, CA 94720}




\begin{abstract}
The stellar initial mass function (IMF), along with the star formation rate, is one of the fundamental properties that any theory of star formation must explain. An interesting feature of the IMF is that it appears to be remarkably universal across a wide range of environments. Particularly, there appears to be little variation in either the characteristic mass of the IMF or its high-mass tail between clusters with different metallicities. Previous attempts to understand this apparent independence of metallicity have not accounted for radiation feedback from high-mass protostars, which can dominate the energy balance of the gas in star-forming regions. We extend this work, showing that the fragmentation of molecular gas should depend only weakly on the amount of dust present, even when the primary heating source is radiation from massive protostars. First, we report a series of core collapse simulations using the ORION AMR code that systematically vary the dust opacity and show explicitly that this has little effect on the temperature or fragmentation of the gas. Then, we provide an analytic argument for why the IMF varies so little in observed star clusters, even as the metallicity varies by a factor of 100. 
\end{abstract}


\section{Introduction}
\label{intro}
The stellar initial mass function - the distribution of stellar masses at birth - appears to be remarkably constant accross a wide range of star-forming environments. While there is some evidence that the IMF is different in giant elliptical \citep{2010Natur.468..940V, 2010ApJ...709.1195T} and dwarf galaxies \citep{2009ApJ...706..599L}, in our own galaxy the IMF is practically universal \citep{2002Sci...295...82K,2003PASP..115..763C,2010ARA&A..48..339B}. In particular, while zero-metallicity stars likely had a much different mass distribution than present-day stars \citep{2002Sci...295...93A, 2002ApJ...564...23B}, the IMFs observed for population I and II stars appear to vary little with the metallicity of the star-forming region. \cite{2010ARA&A..48..339B} review the literature on this topic and find no evidence for a systemic dependence of the IMF on metallicity. For example, in the Milky Way disk, extreme outer galaxy clusters with metallicities of $ \sim 0.1$ solar have the same IMF as nearby clusters with approximately solar metallicity \citep{2008ApJ...675..443Y}. Likewise, globular clusters, which have metallicities that are often 10-100 times lower than the solar value, have the same IMF as regions of present-day star formation once dynamical evolution is taken into account \citep{2000ApJ...530..342D,2010ApJ...718..105D}. Outside our own galaxy, \cite{2008ApJ...681..290S} and \cite{2002ApJ...579..275S} find no difference between the Milky Way disk IMF and that of the Small Magellenic Cloud, which has a metallicity of $\sim 0.2$ solar. In short, the IMF appears to be insensitive to metallicity across a wide range galactic environments, and understanding why is an important challenge for theories of star formation. 

One potential explanation was offered by \cite{2008ApJ...681..365E}. They focus on the characteristic mass of the IMF plateau, which should be related to the Jeans mass $M_J \propto T^{3/2} \rho^{-1/2}$ of the region. They calculate $M_J$ for a prestellar core under the assumption that $T$ is set by the balance between molecular cooling and heating from gas-dust coupling, finding that for a given dust temperature, the gas temperature depends only weakly on metallicity. However, this skirts an important question: what if the dust temperature itself is a strong function of metallicity? In regions where stars are already forming, radiative feedback should dominate the dust's energy balance even before the onset of nuclear burning, especially when the stars are massive. This is not a minor effect. Most stars are born in clusters \citep{2003ARA&A..41...57L}, and the cluster mass function is $dN / dM \propto M^{-2}$ \citep{1999ApJ...527L..81Z,2009ApJ...704..453F,2010ApJ...711.1263C}, which implies equal mass per logarithmic bin. Thus, 2/3 to 3/4 of stars are born in clusters of 1000 $M_{\odot}$ or more, and essentially all of these clusters are expected to contain at least one early O star. Hence, the majority of stars do not form in isolated environments like \citeauthor{2008ApJ...681..365E} consider, but form rather in the presence of a massive star that will affect the gas temperature distribution. If this heating were strongly metallicity dependent, then $M_J$ would be as well.  

Other work has considered the effect of protostellar feedback on the form and apparent universality of the IMF.  \cite{2009MNRAS.392.1363B} argued that the lack of variation in the IMF is the result of self-regulating feedback from radiating protostars, but did not explain why this effect should be independent of metallicity when the material around the stars is optically thick. Additionally, while \citeauthor{2009MNRAS.392.1363B} includes radiative feedback, he underestimates the luminosity by a factor of 20 and the temperature in the core by more than a factor of 2 \citep{2009ApJ...703..131O}. He also does not form any massive stars, which if present would dominate the heating rate. 

High-mass stars are known to strongly affect the temperature and fragmentation of molecular gas. Heating from their extremely high luminosities, which can reach $10^5$ times the solar value, is capable of raising $M_J$ significantly and preventing further fragmentation. Observations using ground-based interferometers reveal that a number of $\sim 100 M_{\odot}$ massive cores appear to remain single, compact objects when observed at 1700 AU resolution, even though the cores initially contained roughly $10^3$ Jeans masses worth of material \citep{2010A&A...518L..85B}. Similarly, \cite{2011ApJ...726...97L} show observations of a young, massive star-forming region within the IRAS 18032-2137 complex that suggest a typical fragment size of $> 1 M_{\odot}$. Numerical simulations (\citealt{2010ApJ...713.1120K}, hereafter KCKM10) find that this effect is most pronounced in regions of high surface density, where the accretion rates are high and the core is effective at trapping radiation in the optically thick regions near protostars. \cite{2008Natur.451.1082K} argued for an effective threshold surface density above which a region will fragment into a few massive objects rather than a few small ones. This potentially explains why the IMF corresponds so well to the core mass function \citep[e.g.,][]{2007A&A...462L..17A,2008ApJ...684.1240E,2004Sci...303.1167B} even for massive ($\gtrsim$ 10 $M_{\odot}$) cores, which based on isothermal assumptions should fragment into many small objects instead of a few massive ones. 

However, the strength of this effect could in principle depend on the opacity of the dust, since that determines the matter-radiation coupling, and hence on the metallicity of the region. It is not obvious that the above effect would work at all in globular clusters, which have far fewer metals, and presumably far less dust, than present-day Milky Way star-forming regions. The purpose of this paper is to gauge the importance of metallicity to the fragmentation of star-forming molecular gas at small scales, where the collapse is highly non-isothermal due to radiative feedback. To do this, we have conducted a series of core collapse simulations where we vary the dust opacity and show that it makes little difference to the core's temperature or its fragmentation. We also provide a simple analytic argument based on the work of \cite{2005ApJ...631..792C} for why the temperature profiles of prestellar cores should depend only weakly on the dust opacity. This work sheds light on why the IMF should be so similar in regions with different metallicity, even when radiation feedback from massive stars on the gas cannot be ignored.

The outline of the paper is as follows. In Section \ref{method} we describe our simulation setup, including the initial conditions and numerical methods used. In Section \ref{results} we show the results of our simulations, which demonstrate how fragmentation is insensitive to metallicity variations of a factor of 20. In Section \ref{discussion} we apply the work of \cite{2005ApJ...631..792C} to show that the temperature profile of a dusty, centrally-heated core should depend only weakly on its metal content. Finally, we present our conclusions in Section \ref{conclusions}.  

\section{Simulation Setup}
\label{method}

\subsection{Numerical Techniques}
To perform our simulations, we have used ORION \citep{1998ApJ...495..821T,1999JCoAM.109..123K}, a parallel, adaptive mesh, radiation-hydrodynamics code for astrophysical applications. The equations and methods used are almost identical to those in KCKM10, so we describe them only briefly here, emphasizing the differences. In short, ORION tracks the four conserved quantities of the combined gas-radiation fluid: the density $\rho$, the momentum per unit volume $\rho v$, the non-gravitational gas energy density $\rho e$, and the radiation energy density $E$. It updates these quantities conservatively, including the effects of gravity and diffuse radiation, but not magnetic fields or ionizing radiation. The radiation transport is solved in the flux-limited diffusion approximation using the mixed-frame formulation \citep{1982JCoPh..46...97M}, retaining terms of order $v/c$. For a full description of the equations solved by the radiation module as well as accuracy tests of the mixed-frame treatment, see \cite{2007ApJ...667..626K} and \cite{Shestakov:2008:MDS:1327548.1327887}. 

In addition to the gas and radiation, the simulation domain also includes star particles, which are placed on fine level grids whenever the Jeans density in a cell exceeds a critical value (see below). Star particles interact with the gas by accreting mass from the simulation volume according to the algorithms in \cite{2004ApJ...611..399K} and emitting radiation according to the protostellar model described in the appendices of \cite{2009ApJ...703..131O}. The full set of equations solved for the fluid is  
\begin{eqnarray}
\label{masscons}
\frac{\partial}{\partial t}\rho & = & - \nabla\cdot(\rho\vecv) - \sum_i \dot{M}_i W(\vecx-\vecx_i), \\
\frac{\partial}{\partial t}(\rho \vecv) & = & -\nabla\cdot(\rho \vecv\vecv) - \nabla P - \rho \nabla \phi - \lambda \nabla E
\nonumber \\
& & {} - \sum_i \dot{\vecp}_i W(\vecx-\vecx_i),
\label{momcons}
\\
\frac{\partial}{\partial t}(\rho e) & = & -\nabla \cdot [(\rho e+P)\vecv] - \rho \vecv \cdot \nabla \phi - \kappa_{\rm 0P} \rho (4 \pi B - c E) 
\nonumber \\
& & + \lambda\left(2 \frac{\kappa_{\rm 0P}}{\kappa_{\rm 0R}} - 1\right) \vecv \cdot \nabla E - \sum_i \dot{\mathcal{E}}_i W(\vecx - \vecx_i),
\label{econsgas}
\\
\frac{\partial}{\partial t}E & = & \nabla \cdot \left(\frac{c\lambda}{\kappa_{\rm 0R} \rho} \nabla E\right) + \kappa_{\rm 0P} \rho (4 \pi B - c E) 
\nonumber \\
& & {} - \lambda \left(2\frac{\kappa_{\rm 0P}}{\kappa_{\rm 0R}} - 1\right) \vecv\cdot \nabla E - \nabla \cdot \left(\frac{3 - R_2}{2} \vecv E\right)
\nonumber \\
& & {}
 + \sum_i L_i W(\vecx - \vecx_i),
\label{econsrad}
\end{eqnarray} 
where $\kappa_{\rm 0R}$ and $\kappa_{\rm 0P}$ are the Planck and Rosseland mean opacities computed in the co-moving frame of the fluid and $B = c a_R T_g^4 / (4\pi)$ is the Planck function. The flux limiter $\lambda$ and $R_2$, which is related to the Eddington factor, are dimensionless numbers that enter into our approximation of the radiative transfer. For details, refer to \cite{2007ApJ...667..626K} and the references therein. 

In the above equations, the sums are taken over all the particles in the simulation. $L_i$ is the luminosity of star $i$ , while $\dot{M}_i$, $\dot{\vecp}_i$, and $\dot{\mathcal{E}}_i$ are the rates at which mass, momentum, and energy are transferred from gas to stars. $W(\vecx-\vecx_i)$ is a kernel that distributes the transfer over a radius of 4 fine level cells around the star. The star particles are updated according to
\begin{eqnarray}
\label{starmass}
\frac{d}{dt} M_i &= & \dot{M}_i, \\
\label{starpos}
\frac{d}{dt} \vecx_i & = & \frac{\vecp_i}{M_i}, \\
\label{starmom}
\frac{d}{dt} \vecp_i & = & -M_i \nabla \phi + \dot{\vecp}_i,
\end{eqnarray}
where $\phi$ is the gravitational potential given by
\begin{equation}
\label{poisson}
\nabla^2\phi = -4\pi G \left[ \rho + \sum_i M_i \delta(\vecx-\vecx_i)\right].
\end{equation} For the purpose of computing the gravitational force on a sink particle, we have used a Plummer Law with a smoothing length of 2 fine level cells (14 AU in physical units) for the gas-sink force and a smoothing length of 1 fine level cell for the sink-sink forces; see \cite{2004ApJ...611..399K} for details. Note that we do not include the effects of protostellar outflows; for discussion of high-mass star formation with outflows, see \cite{Cunningham}. 

We adopt a polytropic equation of state:
\begin{equation}
P = \frac{\rho k_B T_g}{\mu m_{\rm H}} = (\gamma-1) \rho \left(e - \frac{v^2}{2}\right),
\end{equation}
where $T_g$ is the gas temperature, $\mu = 2.33$ is the mean molecular weight for molecular gas with cosmic abundances, and $\gamma$ is the ratio of specific heats. We have taken $\gamma=5/3$, appropriate for a gas of molecular hydrogen with the rotational levels frozen out, but this choice is essentially irrelevant because $T_g$ is determined primarily by radiative effects. 

The above is all identical to KCKM10. The only differences between the numerical schemes employed in those simulations and ours are:
\begin{enumerate}
\item We have used the dust opacity model described in \cite{Cunningham}, based on the work of \cite{2003A&A...410..611S}. This opacity model was created with stellar winds in mind and includes line cooling effects at high temperatures. These effects are irrelevant at $T_g \lesssim 1000$ K, which includes practically all of the gas under discussion here. 
\item We have turned off mergers between sink particles with masses greater than $0.05$ $M_\odot$.
\item We have modified the Plank and Rosseland mean dust opacities by a multiplicative constant $\delta$ to allow for dust-to-gas mass ratios other than solar.  
\end{enumerate}

Item (2) requires some discussion. As mentioned above, each sink particle is surrounded by an accretion zone of 4 fine level cells from which it draws gas. In KCKM10, if a sink particle moved within another sink's accretion zone, the particles would merge regardless of their masses. Because the finest resolution in our simulations is $\sim 7$ AU, this would mean that any stars that ever moved within a distance of 28 AU would be combined. Because this may not be realistic, we have imposed a mass limit of $0.05$ $M_\odot$ above which sink particles will no longer merge. This limit is chosen to roughly correspond to the mass at which a protostar's core temperature becomes high enough to dissociate molecular hydrogen and initiate second collapse \citep{1998ApJ...495..346M,2000ApJ...531..350M}. Before this, the sinks are more like extended balls of gas with radii of a few AU than stars, so it is more likely that they will merge. After second collapse, the sinks represent objects that are much smaller, and mergers should be less likely. Although our initial conditions are also slightly different (see below), this choice accounts for most of the differences between the simulations reported here and those in KCKM10. 
 
\subsection{Refinement criteria}

The computational domain is a cube of side $L_{\text{box}}$ that is discretized into a coarse grid of $N_0$ cells, so that the resolution on the coarse grid is $x_0 = L_{\text{box}} / N_0$. The AMR functionality of ORION automatically identifies regions that need more resolution and covers them with a finer grid. With $L$ levels of refinement and a refinement ratio of 2, the resolution of the finest level is $\Delta x_L$ is $x_0 / 2^L$.  In this work, we have chosen these parameters such that $\Delta x_L$ is always $\sim 7$ AU. 

In generating the grids, a cell is tagged for refinement if
\begin{enumerate}
\item It is within a distance $16\Delta x_L$ of the nearest star particle.
\item It has a density greater than the Jeans density, given by
\begin{equation}
\rho_J = J^2 \frac{\pi c_s^2}{G \Delta x_L^2},
\end{equation}
where $c_s$ is the sound speed and we use $J=1/8$ \citep{1997ApJ...489L.179T}.
\item the local gradient in the radiation energy is greater than a critical value, given by
\begin{equation}
0.15 \frac{E}{\Delta x_L}.
\end{equation}
\end{enumerate} This procedure is repeated recursively until the final level is reached. Taken together, these conditions ensure that the regions near the star particles are always tracked with the highest available numerical resolution. 

\subsection{Initial conditions}

\begin{deluxetable*}{ccccccccccccc}
\tablecaption{Simulation Parameters \label{initial}}
\tablehead{
\colhead{Name} &
\colhead{$\delta$} &
\colhead{$M$ $(\msun)$} &
\colhead{$\Sigma$ (g cm$^{-2}$)} & 
\colhead{$R$ (pc)} &
\colhead{$\sigma_v$ (km s$^{-1}$)} &
\colhead{$t_{\rm ff}$ (kyr)} &
\colhead{$L_{\rm box}$ (pc)} &
\colhead{$L$} &
\colhead{$N_0$} &
\colhead{$\Delta x_0$ (AU)} &
\colhead{$N_L$} & 
\colhead{$\Delta x_L$ (AU)}
}
\startdata
Solar & 1.0 & 300 & 2.0 & 0.1 & 3.59 & 30.2 & 0.40 & 6 & 192 & 430 & 12,288 & 6.71 \\
0.2 Solar & 0.2 & 300 & 2.0 & 0.1 & 3.59 & 30.2 & 0.40 & 6 & 192 & 430 & 12,288 & 6.71 \\
0.05 Solar & 0.05 & 300 & 2.0 & 0.1 & 3.59 & 30.2 & 0.40 & 6 & 192 & 430 & 12,288 & 6.71 \\
High $\Sigma$ &0.2 & 300 & 10.0 & 0.045 & 5.37 & 9.03 & 0.18 & 5 & 168 & 220 & 5,376 & 6.85 \\
\enddata
\tablecomments{Col.\ 7: linear size of computational domain. Col.\ 8: maximum refinement level. Col.\ 9: number of cells per linear dimension on the coarsest level. Col.\ 10: linear cell size on the coarsest level. Col.\ 11-12: same as col.\ 9-10, but for the finest level.
}
\end{deluxetable*}

Our initial conditions also follow the approach laid out in KCKM10, and have been chosen to resemble the structures from which massive stars are believed to form. Observations of the internal structure of infrared dark clouds (IRDCs) using sub-mm interferometry reveal the presence of peaks in the local density distribution termed massive cores \citep{2009ApJ...705.1456S}. They are measured to have masses in the range of $\sim 100 M_{\odot}$, radii of about $\sim$ 0.1 pc, and temperatures of about $\sim 20$ K. Massive cores are observed to be centrally concentrated, and unlike low-mass prestellar cores are highly turbulent, with virial ratios of order unity.

To model these objects, we initialize the simulation volume to contain a sphere of gas with radius $R$, total mass $M$, surface density $\Sigma = M / \pi R^2$, and constant initial temperature $T_{\text{core}} = 20$ K. Following the theoretical models of \cite{2003ApJ...585..850M}, we give the core a power law density profile of  
\begin{equation}
\label{denprof}
\rho(r) \propto r^{-k_{\rho}}
\end{equation} where we have taken $k_{\rho}$ to be 1.5. This is consistent with observations of clumps of molecular gas - structures a few pc in size with thousands of solar masses of material -  which reveal power-law density profiles with scaling exponents between 1 and 2 (\cite{2007A&A...466.1065B}, \cite{1995ApJ...446..665C}, \cite{2002ApJS..143..469M}), and with higher resolution observations of individual cores, which find power law density profiles with slopes between 1.5 and 2 \citep{2011ApJ...726...97L,2009ApJ...696..268Z}. We stress, however, that the artificiality of the initial conditions, in which the density initially lacks structure and the core is considered in isolation, is a potential source of uncertainty. 

The cores are placed in a cubic volume with $L_{\text{box}} = 4 \times R$, surrounded by a background medium with density $\rho_{\text{bg}} = 0.01 \times \rho_{\text{edge}}$, where $\rho_{\text{edge}}$ is the density at the edge of the initial core:
\begin{equation}
\label{rhoedge}
\rho_{\rm edge} = \left(\frac{3-\krho}{4\pi}\right) \frac{M}{R^3}.
\end{equation} To maintain pressure balance, the temperature of the background gas is set to $T_{\text{bg}} = 100 \times T_{\text{core}} = 2000$ K. The opacity of the ambient gas is set to a tiny value to ensure that it does not interfere with the temperature structure of the collapsing core. 

To mimic the effects of turbulence, we also give the cores an initial Gaussian random velocity field with a power spectrum $P(k) \propto k^{-2}$. We scale each component of the velocity to have a one-dimensional root-mean-squared dispersion equal to the velocity at the surface of a singular polytropic sphere \citep{2003ApJ...585..850M}:
\begin{equation}
\label{dispersion}
\sigma_v = \sqrt{\frac{G M }{R (2 k_{\rho} - 1)}}.
\end{equation} The corresponding virial parameter $\alpha_{\text{vir}} = 5 \sigma_v^2 G M / R$ \citep{1992ApJ...395..140B} is 5 for $k_\rho = 3/2$, so that the turbulent kinetic energy is initially larger than the gravitational potential energy. However, the turbulence decays as the core collapses, so the virial parameter drops. 

Starting from these conditions, we have performed a series of four core collapse simulations, summarized in Table \ref{initial}. For three of the runs, we chose a surface density of $\Sigma = 2$ g cm$^{-2}$, which is comparable to the surface densities observed in massive star-forming regions in the Galaxy. We vary the parameter $\delta$, which represents the dust-to-gas mass ratio relative to that expected at solar composition. The values of $\delta$ = 1.0, 0.2, and 0.05 are representative of nearby star clusters, extreme outer galaxy star clusters, and low-metal globular clusters, respectively. We have also conducted a run at $\Sigma = 10$ g cm$^{-2}$, $\delta = 0.2$, which is characteristic of extragalactic super star clusters. Motivated by the above observations, we have chosen to set the mass $M$ of all our cores to $300 M_\odot$. This is higher than in KCKM10, but it will allow us to form massive stars more quickly and thereby gauge the effect of radiative feedback at less computational expense. Because the mass of the cores is the same in the $\Sigma = 2$ g cm$^{-2}$ and $\Sigma = 10$ g cm$^{-2}$ runs, the High $\Sigma$ run has a slightly smaller radius. In the rest of the paper, we will refer to these runs by the names given in Table \ref{initial}. 

We run all the simulations out to $0.5 t_{\text{ff}}$, where $t_{\text{ff}}$ is the gravitational free-fall time evaluated at the mean density $\bar \rho = 3 M / 4 \pi R^3$: 
\begin{equation}
\label{t_ff}
\tff = \sqrt{\frac{3\pi}{32G\overline{\rho}}}.
\end{equation} By this time, the basic similarity of all the runs has been established. We emphasize that in general, the numerical methods and resolution have been held constant across all the runs, so that any differences between them should be attributable to variation in $\delta$ or $\Sigma$. 

\begin{figure*}[h]
  \plotone{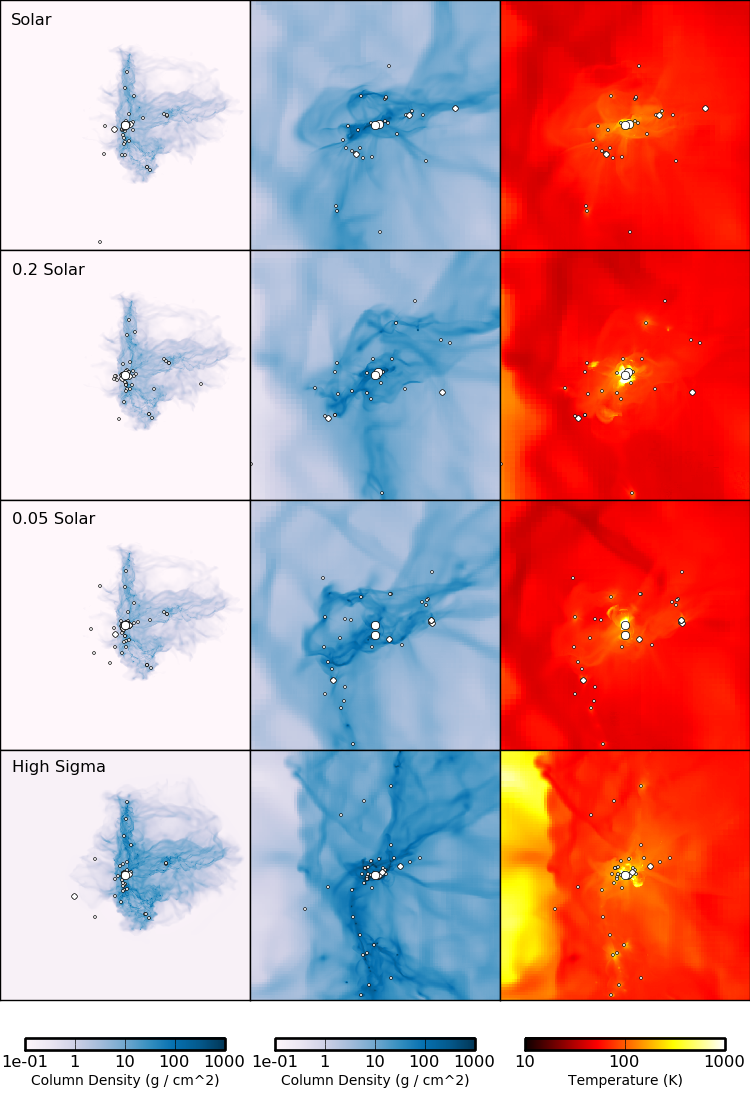}
  \caption{Projections through the simulation volumes at $t = 0.5 t_{\text{ff}}$. The left panels show the column density of the entire core, defined as $\int \rho dx$. The middle column is also the column density, zoomed in to show the middle 5000 AU. The right column shows the column density-weighted temperature, $\int \rho T_{\text{gas}} dx / \int \rho dx$, at the same scale. The rows, from top to bottom, show runs ``Solar," ``0.2 Solar," ``0.05 Solar," and ``High $\Sigma$". Stars are represented by circles drawn on the plots, with the size of the circle corresponding to the size of the star. Stars with masses between $0.05 M_\odot$ and $1 M_\odot$ are the smallest, intermediate mass stars with $1 M_\odot < m < 5 M_\odot$ are larger, and stars with masses greater than $5 M_\odot$ are the largest.}
   \label{panels}
\end{figure*}

\section{Results}
\label{results}
\subsection{Temperature and density structure}

Figure \ref{panels} shows the result of evolving the above cores out to half a global free-fall time. The leftmost panels show the column density in units of g cm$^{-2}$, zoomed out to show the entire simulation volume. The large scale morphology of the collapse is practically identical between the $\Sigma = 2$ g cm$^{-2}$ runs, as that is set primarily by the magnitude of the initial velocity perturbations. The High $\Sigma$ run shows a slight tendency towards more filamentary structure than the others, since the initial Mach number must be larger to pump the higher surface density cloud into virial equilibrium. Nonetheless, the differences between the High $\Sigma$ run and the others are only minor at this scale. 

The middle panels again show the column density, with the same units and color scale as before, but zoomed in to show the central 5000 AU around the most massive object. At this scale, some differences between the runs are apparent. In particular, the shape of the accretion flow around the stars appears to be different in the three $\Sigma = 2$ g cm$^{-2}$ runs. This is because radiation pressure can be important near the massive star(s), and its magnitude varies with the opacity. 

The rightmost panels show the column density-weighted gas temperature at the same 5000 AU scale. The most striking feature is that the temperature of the gas surrounding the central condensation of stars is quite similar in the three $\Sigma = 2$ g cm$^{-2}$ runs. Changing the opacity by a factor of twenty appears to have little impact on the temperature of the bulk of the gas. On the other hand, the High $\Sigma$ run has noticeably more gas heated to a hundred degrees or higher. The ``wall" of hot gas visible on the left-hand side of some of the runs is the hot, diffuse medium that was initially outside the core; it is more visible in the High $\Sigma$ run because the core is smaller, so that the 5000 AU frame captures a larger fraction of the volume. The stars are near the edge of the core at $t = 0.5 t_{\text{ff}}$ in all the runs because the random velocity perturbations we used happened to advect them that way. With a different realization of the turbulence, this would not necessarily happen. 

\subsection{Fragmentation and Star Formation}

\begin{figure}
  \epsscale{1.2}
  \plotone{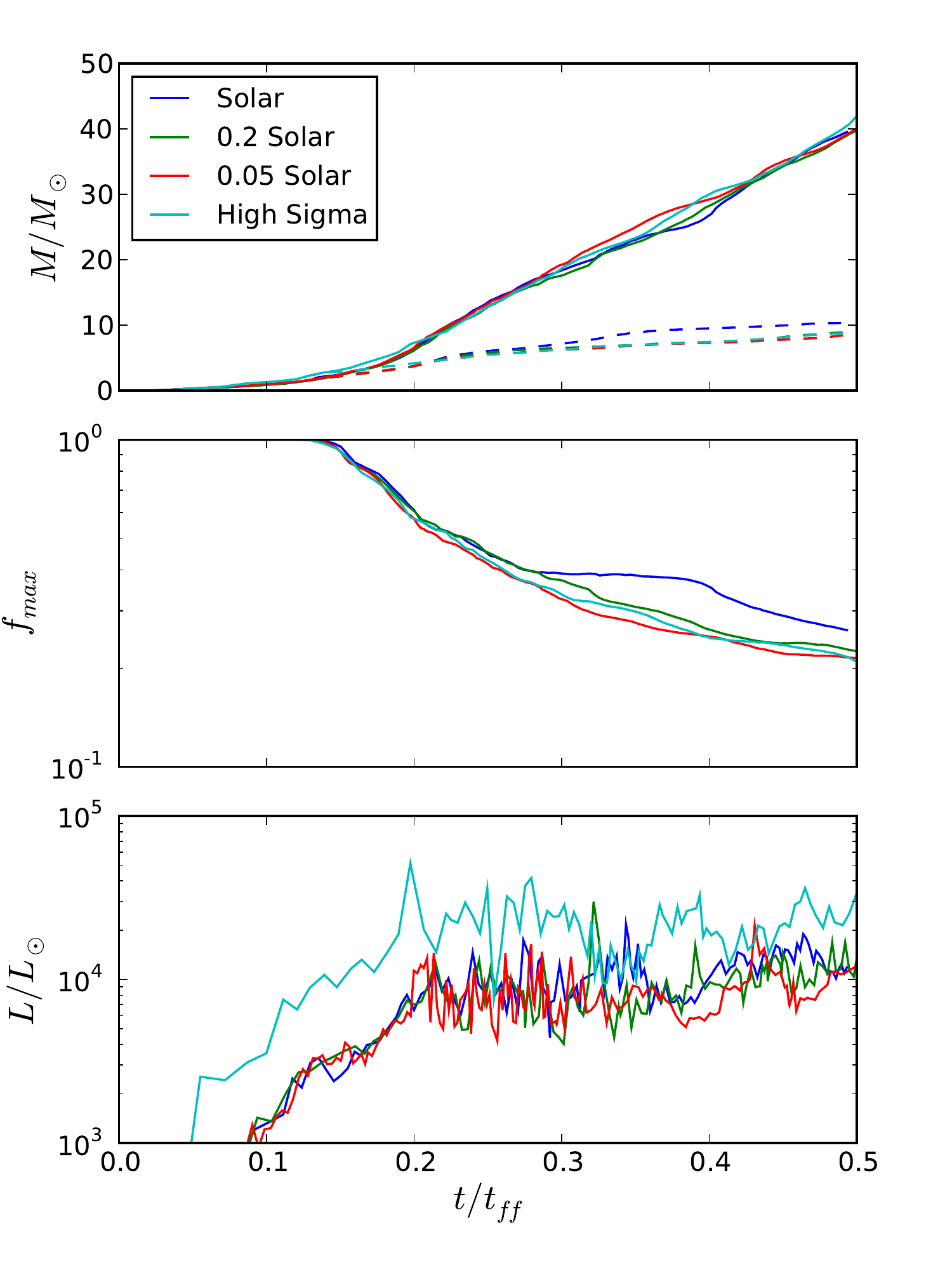}
  \caption{Star particle statistics as a function of time for all four runs. The top panel shows the total mass in stars (the top set of lines) and the mass of the most massive star (bottom, dotted set). The middle plot shows $f_{\text{max}}$, the fraction of total stellar mass that is in the most massive star. The bottom plot is the total stellar luminosity.  }
  \label{stars}
\end{figure}

At the end of  0.5 free-fall times, about 40 $M_\odot$ of gas has been turned into stars, or about 13\% of the total core. At that point, all the runs have a massive star of $\sim 10 M_\odot$. In the three $\Sigma = 2.0$ g cm$^{-2}$ runs, this star forms a binary system with a massive companion of $\sim 6 M_\odot$. In the High $\Sigma$ run, the secondary is only $3 M_{\odot}$, with the missing mass spread out among the other objects.   

A major difference between these simulations and those of KCKM10 and \cite{Cunningham}, is that we form many more objects during the collapse. This difference is largely due to the choice of merger criterion - by turning off mergers beyond a threshold mass, objects that would have formed a single star here form several. At the opening stage of the collapse, the initial velocity perturbations create a network of dense filaments, visible in the leftmost panels of Figure \ref{panels}. Because the core is centrally concentrated, gas falls into the center, forming a star with about $3 M_{\odot}$ after approximately 0.15 free-fall times. At this time, additional sinks begin to form in the gas that is falling on to the star. Some of these sinks are small enough that they will merge onto the central star, but many of them accrete enough mass that they surpass the $0.05 M_{\odot}$ threshold and become ``stars" in their own right. These objects fall into the center of the core's gravity well and begin to undergo complex N-body interactions with each other. They stay in the center for several orbits before being thrown out. At that time, they have will have grown to approximately 0.1 to 1 $M_{\odot}$, and will take away with them mass that would have merged with the central star(s) in the simulations without merger suppression. This results in the most massive star growing much less rapidly after $\sim 0.2 t_{\text{ff}}$ than in KCKM10 or \cite{Cunningham}. Thus, with mergers suppressed, we form a massive star or binary plus a system of a few dozen low-mass stars, as opposed to having most of the stellar mass in one system. 

The fragmentation observed in these simulations is qualitatively different from previous simulations with ORION that did not suppress mergers for sinks larger than $0.05 M_{\odot}$. While we emphasize that the handling of stellar mergers is artificial in both cases, the large number of massive stars observed to have close companions \citep{2008MNRAS.386..447S} shows that merging all stars with separations less than 28 AU is clearly unrealistic. For that reason, the model used here is probably more accurate than merging stars regardless of their masses. However, it probably errs in the other direction. For example, because we necessarily soften the gravitational interactions of stars with gas on scales smaller than the grid scale, we underestimate the dynamical friction that stars experience as they pass through one another's disks, and thus underestimate the rate at which this process causes stars to be captured \citep[e.g.,][]{2005A&A...437..967P,2006A&A...454..811P}. Even if such captures do occur, our present prescription does not allow the stars to merge. We emphasize that all particle merger prescriptions, whether in our code or in others \cite[e.g.,][]{1995MNRAS.277..362B,2010ApJ...713..269F}, are suspect because they involve subgrid models for hydrodynamic and gravitational interactions on scales that are not resolved in the simulation. The absolute number and masses of stars formed is affected by the choice of sub-grid model, so these results be should interpreted cautiously. However, because the merger criterion is the same for all the runs presented in this paper, any differences in the mass distribution of the objects formed between the four runs is due to changes in  $\Sigma$ and $\delta$, not the merger criterion.

The idea that fragmentation of incoming gas limits the mass supply for massive stars has been discussed before in the literature. Most recently, \cite{2010ApJ...725..134P} coined the term ``fragmentation-induced starvation" to describe the phenomenon where gas en route to a massive star instead collapses to form a low mass star before arriving, depriving the massive star of that material. Because our simulations start from turbulent initial conditions, the fragmentation happens in the dense filaments that feed the central star system rather than in a disk as in the non-turbulent simulations of Peters et. al., but the underlying idea is the same. The competitive accretion models of \cite{1997MNRAS.285..201B} and \cite{2004MNRAS.349..735B} show a similar phenomenon. However, it is important to distinguish between the dynamics in our simulation and those in traditional competitive accretion models. In the absence of radiative feedback, fragmentation always proceeds down to the thermal Jeans mass, $\sim 0.5$ $M_\odot$ \citep[e.g.,][]{2006MNRAS.370..488B}. Only once fragments of that mass form do they then grow to larger masses by Bondi-Hoyle accretion of gas that is not initially bound to the star. The dynamics in our simulation are different, in that the radiatively heated gas in our simulations does not always fragment down to such small masses, and our central massive stars are built by a direct collapse and not by Bondi-Hoyle accretion. The difference becomes apparent if one compares our simulations to those of \cite{2005MNRAS.360....2D}, who start with initial conditions very similar to ours, but do not include radiative feedback. In their simulations, after $0.5 t_{\rm ff}$ they find no stars larger than $\sim 1$ $\msun$ because all the gas has fragmented to small masses, while in our simulations we have $10$ $M_\odot$ stars after a similar time.

Figure \ref{stars} shows the properties of the star particles formed in the simulations as a function of time. To show all the runs on the same plot, we have normalized the time by $t_{\text{ff}}$, but in physical units the High $\Sigma$ simulation evolves about 3 times faster than the others. The top panel shows the total mass in stars, which is quite similar across all the runs. This is to be expected, because the star formation rate is set by the global properties of the flow, which are essentially independent of radiative effects. The middle panel shows $f_{\text{max}}$, the fraction of the total stellar mass that is in the most massive star. There is a period around $0.3 t_{\text{ff}}$ where the Solar run levels off, but this appears to be a temporary phenomenon. By $0.5 t_{\text{ff}}$, $f_{\max}$ is very similar across all the runs. Overall, there appears to be little difference between the runs, either in the total mass converted to stars or in the fraction of that mass that ends up in the most massive star. The bottom panel of Figure \ref{stars} shows the total luminosity of all the stars in the simulation. The value of $10^4$ $L_{\odot}$ we find is typical of observed massive protostars (e.g. \cite{2007prpl.conf..197C}). Here, we can see a difference between the High $\Sigma$ run and the others - because of the higher accretion rates, the total luminosity is higher in the High $\Sigma$ run, although the difference decreases with time as the stars become more massive and more of the radiant output comes in the form of nuclear luminosity. The nuclear luminosity is never dominant, however, and thus the luminosity as a function of time is roughly flat after about $0.2 t_{\text{ff}}$, even though we are forming more stars and the stars are growing more massive. We will make use of the luminosity averaged after over $0.2 t_{\text{ff}}$ to $0.5 t_{\text{ff}}$ in Section \ref{discussion} below.   

Figure \ref{cmfs} shows the cumulative mass distribution of the stars in the four runs - that is, for each value of $m$, the y-axis shows the fraction of the total stellar mass than in is stars with masses greater than $m.$ Visually, there appears to be little difference between the curves, particularly for the $\Sigma = 2$ g cm$^{-2}$ runs. To test whether the initial mass function is indeed the same in the different runs, we have performed two-sided Kolmogorov-Smirnov (K-S) tests between each pair of distributions. The results are shown in Table \ref{ks}. At the 10\% level, we cannot reject the null hypothesis that the three $\Sigma = 2$ g cm$^{-2}$ have the same underlying initial mass function. The High $\Sigma$ distribution, on the other had, does seem to be statistically different from the others. 

\begin{deluxetable}{ccccc}

\tablecaption{p-values from two-sided K-S tests for the simulated mass distributions. \label{ks}}

\tablehead{\colhead{Run} & \colhead{Solar} & \colhead{0.2 Solar} & \colhead{0.05 Solar} & \colhead{High $\Sigma$} \\ 
\colhead{} & \colhead{} & \colhead{} & \colhead{} & \colhead{} } 

\startdata
Solar &  -- &  0.1 &  0.62 &  0.02 \\
0.2 Solar &   &  -- &  0.16 &  0.01  \\
0.05 Solar &   &   &  -- &  0.02 \\
\enddata

\end{deluxetable}


\begin{figure}
  \epsscale{1.2}
  \plotone{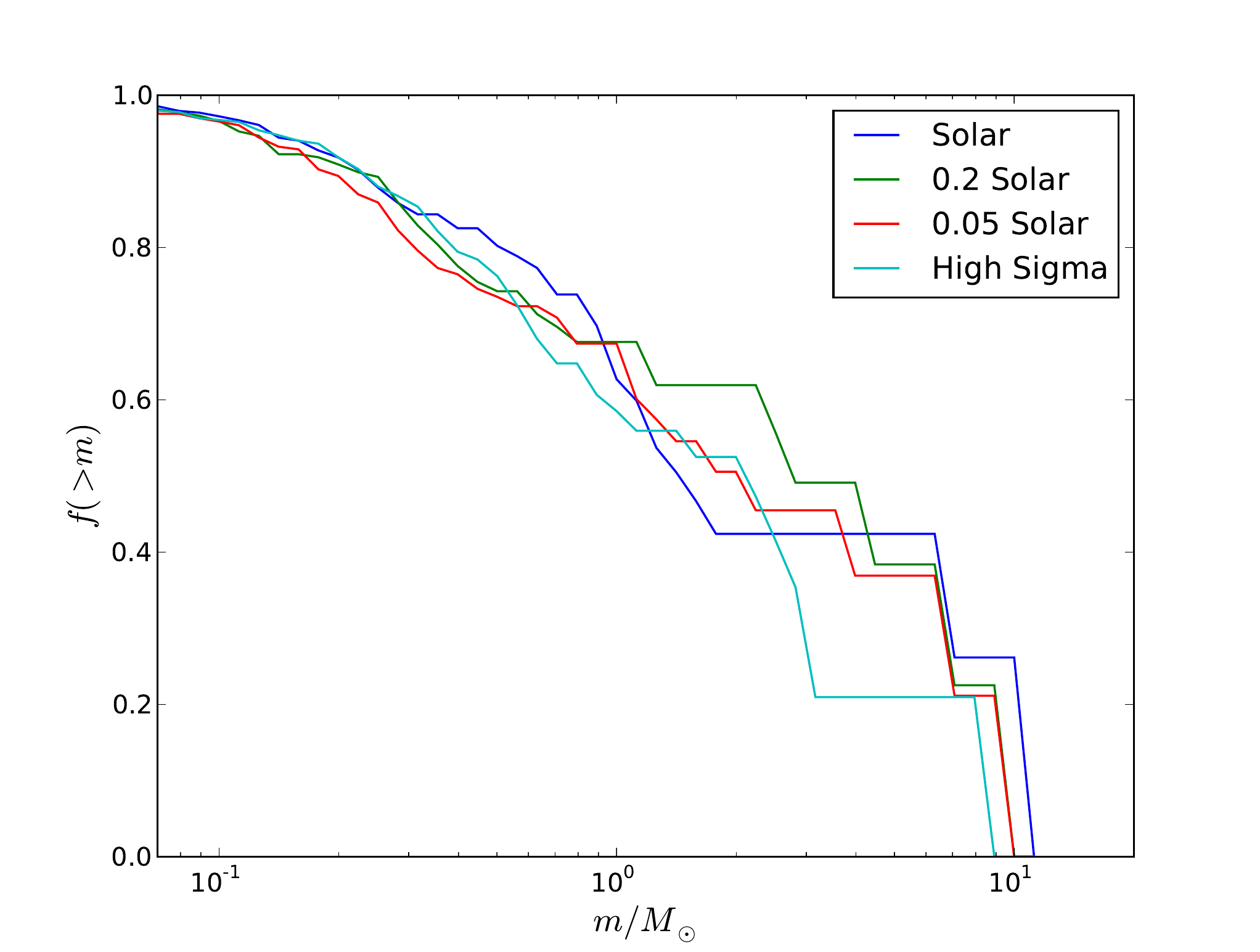}
  \caption{Cumulative mass distributions from all four runs at $t = t_{\text{ff}} 0.5$. $f(>m)$ the fraction of the total stellar mass that is in stars with masses greater than $m.$}
  \label{cmfs}
\end{figure} 

In contrast to the minor effect reported here, \cite{2007ApJ...656..959K} found that isothermal runs fragmented completely differently from radiative ones, and KCKM10 found a major difference in fragmentation between runs with low and high surface density. To summarize, the differences in the fragmentation of all of our runs are minor. To the extent that there are significant differences, they are due to changes in the surface density, rather than to changes in the metallicity. At least within the range of parameters considered here, metallicity appears to have little effect on either the temperature or the fragmentation of molecular gas. 

\section{Discussion}
\label{discussion}
\subsection{Analytic Model}
The above simulations suggest that metallicity plays little role in the fragmentation of star-forming gas. To understand why, consider a simple model system like the initial conditions above: a core of gas and dust with radius $R_c$, mass $M$, surface density $\Sigma = M / \pi R_c^2$, and a power law density profile $\rho(r) \propto r^{-k_\rho}$. We would like to understand what happens to the temperature of this core once stars have started to form, so we will place a point source of luminosity $L$ in the center to represent the combined radiant output of the central collection of stars. We assume that the dust opacity follows a power law in the far IR regime: 
\begin{eqnarray}
\kappa_\nu &=& \delta \kappa_0 \left( \frac{\lambda_0}{\lambda} \right)^\beta \nonumber \\
&=& \delta \kappa_0 \left( \frac{T}{T_0} \right)^\beta, \text{ for 3 mm} < \lambda < 30 \text{ } \mu \text{m}
\end{eqnarray} where the subscript ``$0$" refers to an arbitrary reference value and $\delta$ is the dust-to-gas mass ratio relative to solar. The $T$ in this equation is the dust temperature, which we assume is identical to the radiation temperature.  We will adopt the dust opacity model of \cite{2001ApJ...548..296W}, for which $\kappa_0 = 0.27$ cm$^2$ g$^{-1}$ at $\lambda_0 = 100$ $ \mu \text{m}$ and $\beta = 2$; however, we have verified that using the opacities of \cite{2003A&A...410..611S} (as used by ORION) or \cite{1994ApJ...421..615P} makes little difference. For reference, the corresponding $T_0$ is 144 K.

The emission from such a system is considered in \cite{2005ApJ...631..792C} (hereafter CM2005). They find that even though cores do not have sharply delineated photospheres as stars do, the radiation they emit is still well-described by 
\begin{equation}
L = \tilde{L} 4 \pi R_{\text{ch}}^2 \sigma T_{\text{ch}}^4 
\end{equation} where $\tilde{L}$ is a constant of order unity and $T_{\text{ch}}$ and $R_{\text{ch}}$ are a characteristic temperature and the radius from which radiation with frequency $\nu_{\text{ch}} = k T_{\text{ch}} / h$ has an optical depth of 1. These are given by
\begin{eqnarray}
\tilde{R_c} &=& \frac{R_c}{R_{\text{ch}}} \nonumber \\
&=& \left\{\frac{(L/M)\Sigma^{(4+\beta)/\beta}}{4\sigma \tilde{L}}
\left[\frac{(3-k_\rho) \delta \kappa_{0}}{4(k_{\rho}-1)T_0^\beta}\right]
^{4/\beta}\right\}^{-\frac{\scriptstyle \beta}{\scriptstyle \alpha}}\; 
\label{eq:rct}
\end{eqnarray} and 
\begin{eqnarray}
T_{\rm ch}=\left\{\frac{L/M}{4\sigma \tilde{L} 
\Sigma^{\frac{3-k_{\rho}}{k_{\rho}-1}}}
\left[\frac{4(k_{\rho}-1)T_{0}^{\beta}}{(3-k_{\rho}) \delta \kappa_{0}}\right]
^{\frac{2}{k_{\rho}-1}} \right\}
^{\frac{\scriptstyle k_{\rho}-1}{\scriptstyle \alpha}}\; 
\label{eq:tch}
\end{eqnarray} where $\alpha = 2\beta+4(k_{\rho}-1)$. Rather than using the expression for $\tilde L$ given in CM2005, we will adopt the more accurate expression from \cite{2008ApJ...683..693C}, which they report gives excellent agreement with results from the DUSTY code, based on the work of \cite{1997MNRAS.287..799I}: 
\begin{equation}
\label{Ltilde}
\tilde{L} = 1.6 \tilde{R_c}^{0.1}.
\end{equation} A final result we will take from CM2005 is that the temperature profile in the vicinity of the photosphere is also well-described by a power law:
\begin{equation}
T(r) = T_{\text{ch}} \left( \frac{r}{R_{\text{ch}}} \right)^{-k_T}
\end{equation} We can solve the above equations simultaneously to get that the temperature as a function of density (or, equivalently, radius) in the core is 
 \begin{eqnarray}
 \label{T}
&T(\rho)& = T_c \left(\rho/\rho_{\text{edge}} \right)^{k_T/k_\rho}, \nonumber \\
&T(r)& = T_c \left( r / R_c \right)^{-k_T}, \nonumber \\
&T_c& = \left[ \frac{L/M}{4 \sigma \tilde{L}} \right]^{\frac{k_\rho - 1 + \beta k_T}{\alpha}} \left[ \frac{(3 - k_\rho) \delta \kappa_0}{4 (k_\rho -1) T_0^\beta} \right]^{\frac{4 k_T - 2}{ \alpha}} \nonumber \\
& \times & \Sigma^{\frac{(4 + \beta) k_T + k_\rho - 3}{\alpha}},
\end{eqnarray} or, specializing to $k_\rho = 3/2$ and $\beta = 2$, 
 \begin{eqnarray}
T_c &=& \left[ \frac{L/M}{4 \sigma \tilde{L}} \right]^{\frac{k_T + 1/4}{3}} \left[  \frac{3 \delta \kappa_0}{4 T_0^\beta} \right]^{\frac{2 k_T - 1}{3}} \Sigma^{k_T - 1/4}.
\end{eqnarray} 

We can see from this expression that the scaling of $T_c$ with $\delta$ actually goes to zero for $k_T = 0.5$. This is surprising at first, since both $\tilde{R_c}$ and $T_{\text{ch}}$ scale with $\delta$, to the -2/3 and -1/3 powers, respectively, for our assumed parameter values. However, we can understand this in the following way: take a point $r$ that is outside the effective photosphere of the core. Since radiation with frequency $\nu_{\text{ch}}$ is thin here, this point's temperature will be determined roughly by radiative equilibrium with a luminosity source with effective radius $R_{\text{ch}}$ and temperature $T_{\text{ch}}$. If we then lower $\delta$, $R_{\text{ch}}$ will decrease, since the core will be less effective at trapping radiation and photons with frequency $\nu_{\text{ch}}$ will be able to travel further through the core's envelope before being absorbed. This will make the effective emitting area smaller, which will tend to lower the flux at point $r$. However, because it is closer to the central heating source, the temperature at the new value of $R_{\text{ch}}$ will be higher as well, and that will tend to increase the flux at $r$. If the temperature always scales as the -0.5 power of radius, then these two effects will cancel out exactly; the luminosity seen at $r$, $L \approx 4 \pi \sigma R_{\text{ch}}^2 T_{\text{ch}}^4$, will be the same, and the temperature will be the same as well. 

Is $k_T$ close to 0.5? CM2005 give a fitting function for the temperature scaling exponent $k_T$ as a function of $\tilde{R_c}$:
\begin{equation}
\label{eq:k}
k_T = \frac{0.48 k_\rho^{0.005}}{\tilde{R_c}^{0.02 k_\rho^{1.09}}} + \frac{0.1 k_\rho^{5.5}}{\tilde{R_c}^{0.7 k_\rho^{1.09}}}
\end{equation} This expression shows that as long as $\tilde R_c > 1$, meaning that the photospheric radius is less than half the core radius, $k_T$ is indeed quite close to 0.5 and depends only weakly on $R_c$. This expression assumes that $T_{\text{ch}} \lesssim $ 300 K, so that most of the flux is emitted at wavelengths longer than 30 $\mu$m and the opacity is well approximated by a power law in frequency. We have verified that this condition holds under the circumstances considered in this paper. Since dust sublimates at $\sim$ 1000 K, this condition also implies that $R_{\text{ch}}$ is larger than the dust destruction radius. 

Note that for a given $\Sigma$ and $L/M$, $\tilde R_c$ is a function of $\delta$ only. From our simulation results, we can calculate the light-to-mass ratio associated with $\Sigma = 2$ and $10$ g cm$^{-2}$ from the luminosities shown in the bottom panel of Figure \ref{stars}, averaged over the period from $t = 0.2 t_{\text{ff}}$ to $t = 0.5 t_{\text{ff}}$. The results are shown in Table \ref{LtM}. The ``$M$" here refers to the gas mass only, not the mass in stars. In the $\Sigma = 2$ g cm$^{-2}$ runs, the light-to-mass ratio appears to depend weakly on $\delta$, but the effect is small and we will simply use the average over the three metallicities. 

\begin{deluxetable}{cc}

\tablewidth{2.0 in}

\tablecaption{Light-to-Mass Ratios \label{LtM}}

\tablehead{\colhead{Run} & \colhead{$L/M$} \\ 
\colhead{} & \colhead{} } 
\startdata
Solar &  80.7 \\
0.2 Solar &  69.8  \\
0.05 Solar &  65.4 \\
High $\Sigma$ &  152.6 \\
\enddata
\tablecomments{Averaged over $t = 0.2 t_{\text{ff}}$ to $t = 0.5 t_{\text{ff}}$.} 
\end{deluxetable}

\begin{figure}
  \epsscale{1.2}
  \plotone{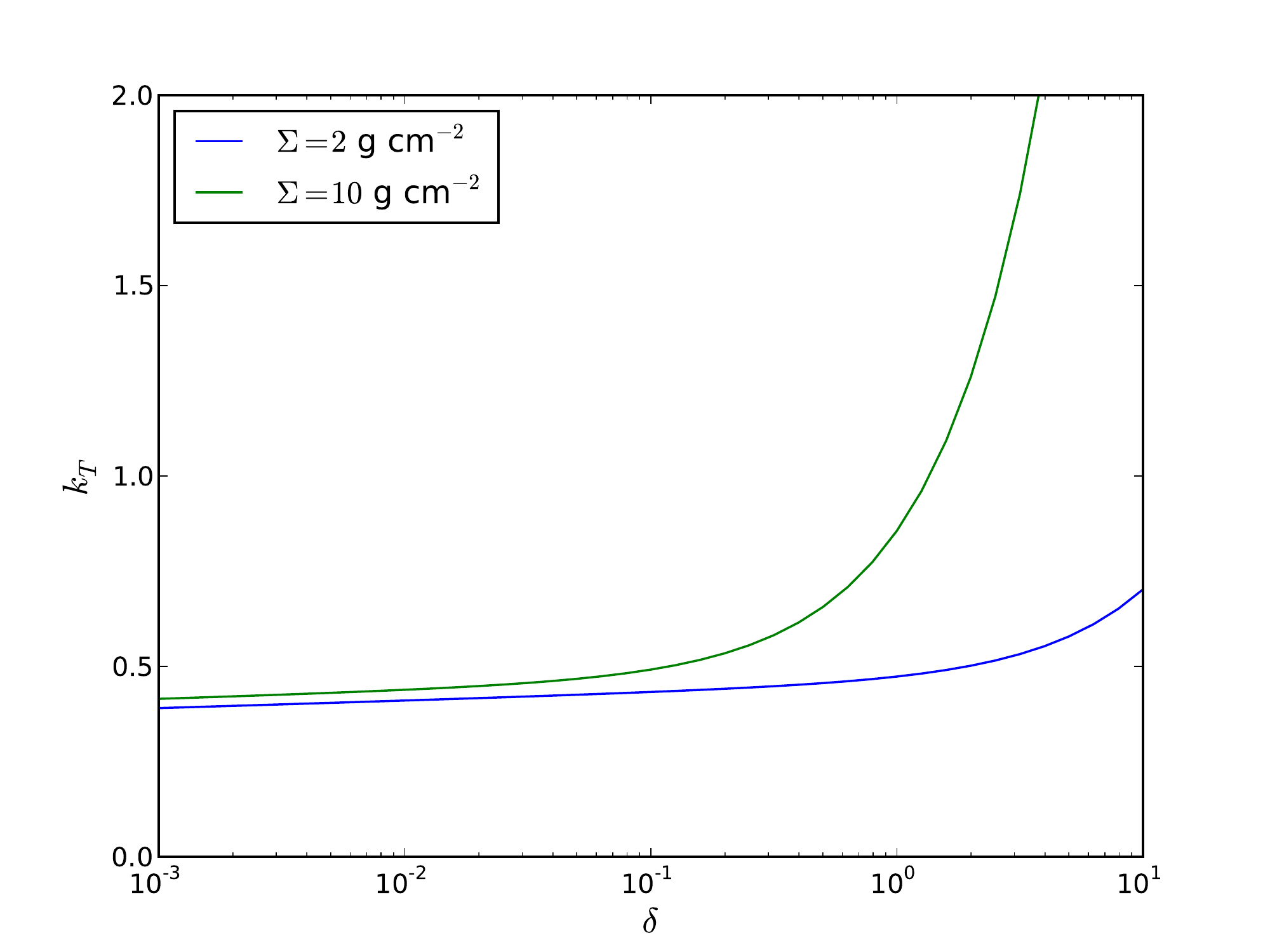}
  \caption{Temperature scaling exponent $k_T$ as a function of $\delta$ for the two different values of $\Sigma$ considered in the simulations.}
  \label{kT}
\end{figure}

In Figure \ref{kT}, we plot $k_T$ as a function of $\delta$ for the two values of $\Sigma$ represented in our simulations from Equation (\ref{eq:k}). Note that $k_T$ remains relatively close to 0.5 for a wide range of values for $\delta$ for both of the surface densities we have considered. It is therefore not surprising that the simulations show such a weak dependence on $\delta$. 

Note that for most of the parameter space, the value of $k_T$ is actually a bit less than 0.5, meaning that the temperature actually scales inversely with $\delta$. When the temperature falls off more slowly than $k_T = 0.5$, the temperature at the smaller value of $R_{\text{ch}}$ associated with a lower $\delta$ will actually be larger than what is required to keep the temperature outside of the photosphere at a constant value, so $T$ will increase slightly. Once $\delta$ increases much past solar, however, $k_T$ increases dramatically and $T$ begins to rise with $\delta$. 

\subsection{Comparison to simulations}

\begin{figure}[h!]
  \epsscale{1.2}
  \centering
    \plotone{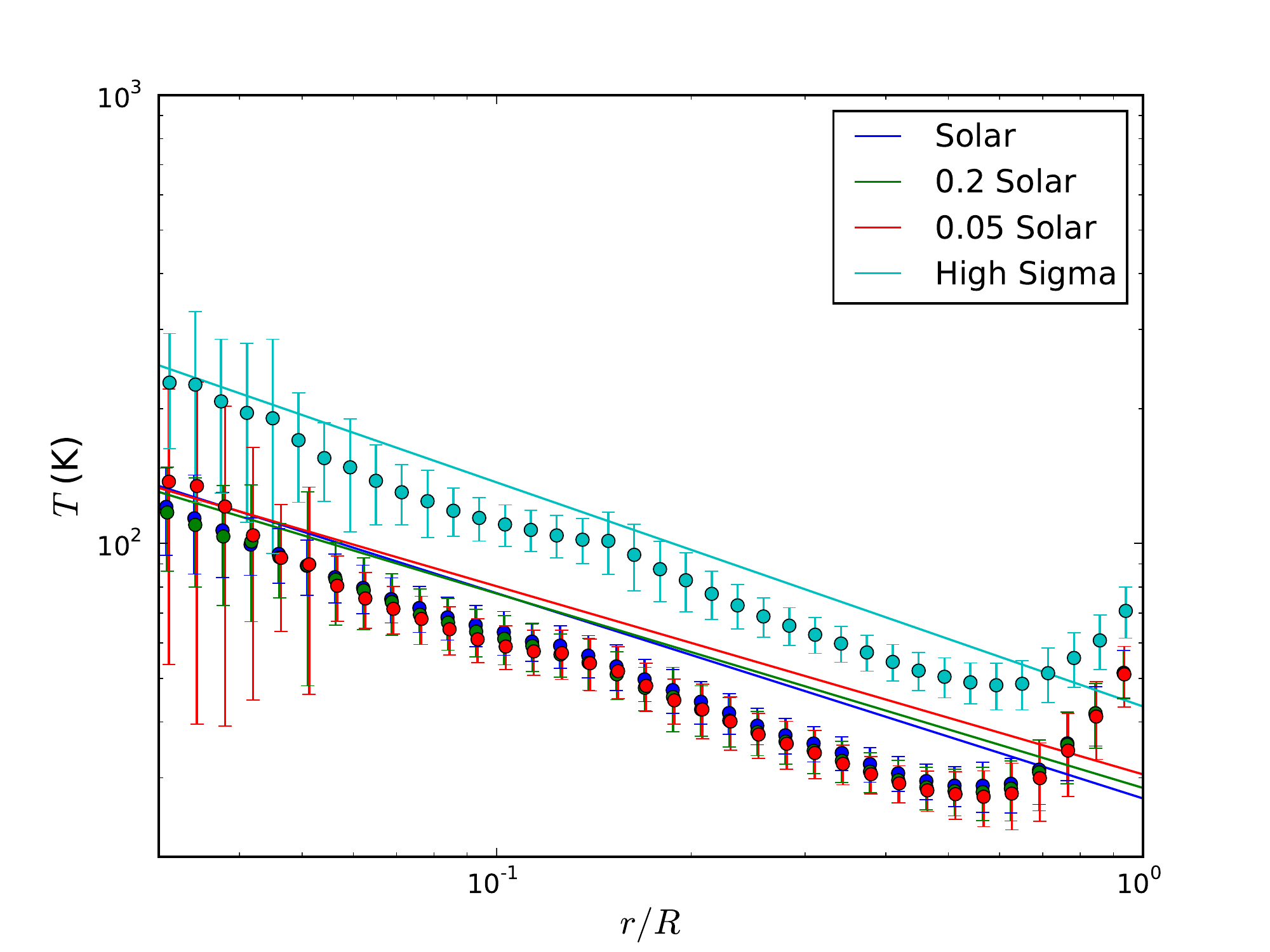}
    \caption{Temperature profiles from the simulations (dots) and analytic theory (solid lines). To show both values of $\Sigma$ on the same plot, we have normalized $r$ by the size of the core $R_c$. The simulations profiles are averaged over $t = 0.2 t_{\text{ff}}$ to $t = 0.5 t_{\text{ff}}$.}
  \label{profiles}
\end{figure}

We can also calculate the temperature profiles expected for the cores in the above simulations. The procedure is as follows. Using the light-to-mass ratios from Table \ref{LtM} along with the known simulation values of $\Sigma$ and $\delta$, we can compute $k_T$ using Equations (\ref{eq:rct}) and (\ref{eq:k}). Then, we use this value along with Equation (\ref{T}) to get the temperature as function of radius. In Figure \ref{profiles}, we plot these relations and compare them to the profiles calculated from the simulations. To get the simulation profiles, we calculated the density-weighted mean temperature in spherical shells of radius $r$ around the most massive star, and plotted the result versus $r$. The density weighting ensures that the hot, diffuse gas surrounding the cores does not interfere with the result inside the core, although its presence can be seen in the temperature rise as $r$ approaches the core radius $R_c$. We have averaged the simulation profiles over all the snapshots from $t = 0.2 t_{\text{ff}}$, when the luminosity has roughly leveled off, to $t = 0.5 t_{\text{ff}}$, and the error bars show the standard error over all the snapshots. The error bars are larger close to the central massive star because of the presence of dynamically interacting stars within in central few hundred AU of the simulation volume. The important things to note are 
\begin{enumerate}
\item The simulation results confirm the power law nature of the temperature profile and agree closely with the predicted slopes
\item The effect of varying the metallicity is quite small compared to the effect of changing the surface density, both in the analytic calculation and in the simulations.
\end{enumerate}

The analytic expression is systematically higher than the simulations by roughly 10\%. It does, however, agree with the jump in $T$ from $\Sigma = 2$ g cm$^{-2}$ to $\Sigma = 10$ g cm$^{-2}$ in quite well.  The discrepancy is likely due to differences in the treatment of the radiative transfer between our simulations and CM2005. ORION treats the radiation in a frequency-averaged gray approximation in terms of the Plank and Rosseland mean opacities, while CM2005 used the DUSTY code \citep{1997MNRAS.287..799I}, which includes frequency information about the photons. However, CM2005 also assumed a spherically symmetric, static core with no density perturbations, so it is not clear which result is more accurate. Whatever the case, both methods agree that the temperature profiles are not particularly sensitive to $\delta$.

\subsection{Predictions for star-forming regions}

Star-forming regions in the Milky Way and other galaxies sample a large range of surface densities and metallicities. A question we can ask is: in what range of parameter space is the gas temperature insensitive to changes in the metallicity? To answer this question, we will use the relationship between the light-to-mass ratio and the surface density given in \cite{2008Natur.451.1082K}:
\begin{equation}
L/M \approx 3.6 M_2^{-0.33} \Sigma_0^{0.67} T_{\text{b,1}}^{0.16} \left( \frac{L_{\odot}}{M_{\odot}} \right)
\end{equation} where $M_2$ is $M / 100 M_\odot $, $\Sigma_0 = \Sigma / 1 \text { g cm}^{-2}$, and $T_{\text{b,1}}$ is the background core temperature divided by 10K. In what follows, we will take $M_2 = 3$ and $T_{\text{b,1}} = 2$. This expression is calculated by assuming that the core converts its gas into stars at a rate of a few percent per free-fall time and is not accurate once massive stars have formed. However, it is still relevant to the early stages of collapse, and will allow us to gauge whether the core temperature becomes high enough to slow fragmentation. Note that the star formation efficiency assumed here is lower than what we see in the simulations; because we have not included the effects of outflows or other turbulence-driving mechanisms, our star formation rates are higher than observed. 

\begin{figure}[h!]
  \epsscale{1.2}
    \centering
    \plotone{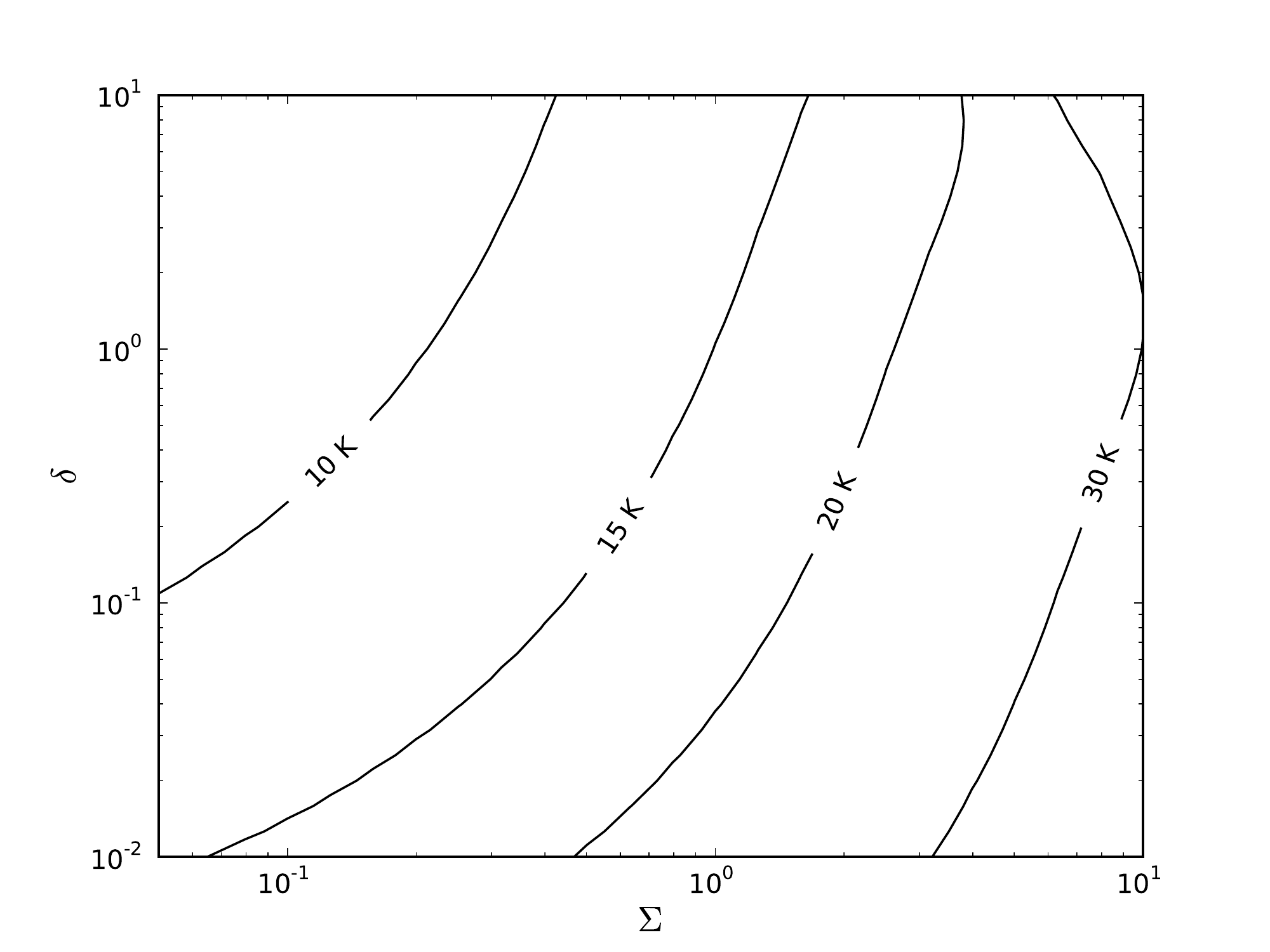}
    \caption{Contours of the mean core temperature $T(\bar \rho)$ at the early stages of collapse as a function of $\Sigma$ and $\delta$ for a core with mass $M = 300 M_{\odot}$. The turnover of the $T = 30 $ K contour at high $\Sigma$ and $\delta$ is due to $k_T$ becoming large; see Equation (\ref{eq:k}) and Figure \ref{kT}.}
    \label{contour}
\end{figure}

Using this expression, we can eliminate the dependence on $L$ in the equation for $T_c$ to get the edge temperature as a function of $M$, $\Sigma$ and $\delta$ only. We plot the expected temperature for a $M = 300 M_{\odot}$ core as a function of these parameters in Figure \ref{contour}. To characterize the core by a single temperature, we use Equation (\ref{T}) to evaluate $T(\rho)$ at the mean density $\bar{\rho} = 3 \rho_{\text{edge}} / (3 - k_\rho)$. Note that we have assumed the temperature is determined by protostellar feedback; gas with sufficiently low metallicity would be warm anyway owing to the lack of coolants. \cite{2010ApJ...722.1793O} considered this effect, and concluded that it should be dominant for metallicities lower than 0.01 $Z_{\odot}$, so we will limit our analysis to values of $\delta$ greater than $0.01$. Figure \ref{contour} includes a range of surface densities that span the conditions typical of star formation, from low-mass star-forming regions like Taurus and Perseus (0.1 g cm$^{-2}$), to regions of active massive star formation in the galaxy (1 g cm$^{-2}$), to extra-galactic super star clusters (10 g cm$^{-2}$). For all those environments, one would need to change $\delta$ by at least two orders of magnitude to get a factor of 2 change in the temperature. Hence, we expect there to be very little variation in the fragmentation of cores across environments with different metallicities over the range currently probed by observations. Note that Figure \ref{contour} shows the mean core temperature only when the dominant source of heating is protostellar feedback. In reality, cores in the upper left hand region of the parameter space would likely be hotter than the indicated 5 - 10 K due to heating from cosmic rays or the decay of turbulence, so the true range of temperature variation is even smaller than indicated in Figure \ref{contour}.

Unlike $\delta$, $\Sigma$ does make a significant difference in the core temperature. In regions like Taurus ($\Sigma \sim 0.1 $ g cm${^-2}$), we do not predict protostellar feedback to be able to heat the gas much above 10 K. Hence, collapse there will likely be isothermal, and may be more prone to fragmentation to produce low-mass stars. Regions like of higher surface density ($\Sigma \sim 1 $ g cm$^{-2}$), are not isothermal, which may bias them towards forming high-mass stars. 

\section{Conclusions}
\label{conclusions}
We have performed a series of numerical experiments using ORION in which we follow the collapse of massive cores past the onset of star formation to the subsequent heating of the gas by radiative feedback, varying the dust opacity by a factor of twenty. We find that the opacity makes little difference to either the temperature or the fragmentation of the cores as they collapse. Our simulations consider surface densities of $\Sigma = 2$ g cm$^{-2}$, characteristic of massive star-forming regions in the Milky Way, and $\Sigma = 10 $ g cm$^{-2}$, characteristic of extra-galactic super star clusters. 

We have also presented an analytic argument for why the IMF should be relatively independent of the metallicity of the star-forming region, even when the heating is dominated by a central source as in high-mass star-forming cores.  This helps to explain why there do not appear to be significant differences between the IMFs of disk stars and globular clusters, or between those of the Milky Way and the SMC, despite large differences in metallicity. We find that the metallicity should only weakly influence the IMF over a large range of star-forming environments. 

\acknowledgements We thank Charles Hansen, Stella Offner, Andrew Cunningham, and the anonymous referee for helpful comments. This project was initiated during the ISIMA 2010 summer program, funded by the NSF CAREER grant 0847477, the France-Berkeley fund, the Institute for Geophysics and Planetary Physics and the Center for Origin, Dynamics and Evolution of Planets. We thank them for their support. Support for this work was also provided by: the DOE SciDAC program under grant DE-FC02-06ER41453-03 (ATM), an Alfred P. Sloan Fellowship (MRK), NSF grants AST-0807739 (MRK), CAREER-0955300 (MRK), and AST-0908553 (CFM, RIK and ATM), the US Dept. of Energy at LLNL under contract DE-AC52-07NA (RIK), NASA through Astrophysics Theory and Fundamental Physics grant NNX09AK31G (RIK, CFM, and MRK), and through a Spitzer Space Telescope Theoretical Research Program grant (MRK and CFM). Support for computer simulations was provided by an LRAC grant from the NSF through Teragrid resources and NASA through a grant from the ATFP. We have used the YT software toolkit \citep{2010arXiv1011.3514T} for data analysis and plotting. 

\bibliographystyle{apj}
\bibliography{bibliography}	

\end{document}